\documentclass[twocolumn,amsmath,amssymb,aps]{revtex4}
\usepackage{graphicx}
\usepackage{dcolumn}
\usepackage{bm}
\usepackage{latexsym,amsmath,amssymb,amsbsy}
\usepackage{graphicx}
\usepackage{nccfoots}
\usepackage[symbol]{footmisc}

\begin{document}


\title{Irradiation of Nuclear Track Emulsions with Thermal Neutrons, \\
Heavy Ions, and Muons}

\author{\textbf{D.~A.~Artemenkov$^{\textbf{1),~{*}}}$, V.~Bradnova$^{\textbf{1)}}$, A.~A.~Zaitsev$^{\textbf{1), 2)}}$, P.~I.~Zarubin$^{\textbf{1)}}$, I.~G.~Zarubina$^{\textbf{1)}}$, R.~R.~Kattabekov$^{\textbf{1), 3)}}$, K.~Z.~Mamatkulov$^{\textbf{1), 4)}}$, V.~V.~Rusakova$^{\textbf{1)}}$}
\Footnotetext{1)}{Joint Institute for Nuclear Research, ul. Joliot-Curie 6, Dubna, Moscow oblast, 141980 Russia.}
\Footnotetext{2)}{Smolensk State University, ul Przheval'skogo 4, 214000 Russia.}
\Footnotetext{3)}{Physical-Technical Institute, Uzbek Academy of Sciences, ul. Mavlyanova 2, Tashkent, 700084 Republic of Uzbekistan.}
\Footnotetext{4)}{A. Kodiriy Jizzakh State Pedagogical Institute, Sharaf Rashidov pr. 4, Jizzakh City, 130100 Republic of Uzbekistan.}
\Footnotetext{*}{E-mail: \texttt{artemenkov@lhe.jinr.ru}}}

\indent \par
\noindent \affiliation{Received XX, 2014}

\begin{abstract} 
\indent Exposures of test samples of nuclear track emulsion were analyzed. Angular and energy correlations of products originating from the thermal-neutron-induced reaction n$_{th} + ^{10}$B $\rightarrow ^{7}$Li $+ (\gamma) + \alpha$ were studied in nuclear tack emulsions enriched in boron. Nuclear track emulsions were also irradiated with $^{86}$Kr$^{+17}$ and $^{132}$Xe$^{+26}$ of energy about 1.2 MeV per nucleon. Measurements of ranges of heavy ions in nuclear track emulsions made it possible to determine their energies on the basis of the SRIM model. The formation of high-multiplicity nuclear stars was observed upon irradiating nuclear track emulsions with ultrarelativistic muons. Kinematical features studied in this exposure of nuclear track emulsions for events of the muon-induced splitting of carbon nuclei to three alpha particles are indicative of the nuclear-diffraction interaction mechanism.\par

\indent \par
\noindent \textbf{DOI:}  10.1134$/$S106377881504002X
\end{abstract}          

\maketitle

\begin{center}
INTRODUCTION
\end{center}
\indent Possessing an excellent sensitivity and spatial resolution, nuclear track emulsions preserve their position as a universal and relatively cheap detector for surveying and searching investigations into nuclear and particle physics. The use of this classic procedure in beams from modern accelerators and reactors proved to be quite successful. In a number of important problems, the completeness of observations that is ensured by nuclear track emulsions remains inaccessible for electronic detection methods. In particular, the clustering of the whole family of light nuclei, including radioactive ones, has been studied over the past decade in the dissociation of relativistic nuclei in nuclear track emulsions [1–4]. In the region of low energies, the decays of $^{8}$He nuclei implanted in nuclear track emulsions [5] and the breakup of $^{12}$C nuclei into three alpha particles under the effect of thermonuclear neutrons [6] were analyzed in recent years.\par
\indent At the present time, there arise new problems associated with the calibration of heavy-ion ranges in nuclear track emulsions. Via solving such problems, it would become possible to extend a methodological basis for studying new aspects of fission physics in the region of heavy nuclei by means of the track-emulsion procedure. The use of automated microscopes makes it possible to approach a new level of application of nuclear track emulsions. The development of nuclear track emulsions characterized by a submicron resolution opens new horizons in searches for hypothetical dark-matter particles by tracks of recoil nuclei. Thus, the return of the emulsion method in the practice of nuclear-physics experiments at the state-of-the-art level provides new possibilities for solving a broad range of problems. The prospects for this method were recently discussed at a dedicated workshop [7].\par
\indent Test track-emulsion samples manufactured by the MICRON production unit of the Slavich Company JSC [9] are being presently irradiated within the Becquerel project [8]. The samples in question are created by casting emulsion layers 50 to 200 $\mu$m thick onto glass substrates. The basic properties of this nuclear track emulsion are close to those of BR-2 nuclear track emulsions, which are sensitive to relativistic particles. The production of BR-2 nuclear track emulsions had been performed for more than four decades and was completed about ten years ago. The product emulsion [9] has already been used in the range-based spectrometry of alpha particles [3, 4].\par
\indent Test irradiations were aimed primarily at a general quality control and a control of the emulsion sensitivity to relativistic particles, as well at a comparison of ranges of slow nuclei that have strongly ionizing low energies with the values calculated on the basis of the SRIM simulation code [10]. Not only do exposures of the product nuclear track emulsion to beams from modern accelerators and reactors permit performing range–energy calibrations, but this also leads to observations and conclusions of value in and of themselves. Also, these exposure give impetus to the development of the track-emulsion method itself since they provide new data for evolving automated microscopes and for extending the range of nuclear-physics education. The present article combines the results obtained by analyzing recent exposures of nuclear track emulsions to thermal neutrons, low-energy heavy ions, and ultrarelativistic muons. So wide a variety of experimental implementations, including those in [3, 4], became possible owing to the use of the new nuclear track emulsion, whose properties permitted applying the same strategy to coordinate measurements for tracks of length between several microns and several tens of microns. The video data on the interactions studied in the track emulsion are available on the website quoted in [8].\par

\begin{center}
EXPOSURE TO THERMAL NEUTRONS
\end{center}

\indent The addition of boracic acid to nuclear track emulsions makes it possible to solve problems of practical importance in beams of thermal neutrons ($n_{th}$) — that is, to determine their profiles and fluxes. Enrichment of track emulsions in boron permits observing charged products of the reaction $n_{th} + ^{10}$B $\rightarrow ^{7}$Li $+ (\gamma) + ^{4}$He. This reaction, which leads to an energy release of 2.8 MeV, proceeds with a probability of about 93\% via the emission of a 478-keV photon by the product $^{7}$Li nucleus from  the only excited state. Track-emulsion samples doped with boracic
acid were irradiated with thermal neutrons nth at an intensity of about $10^{7}n_{th} s^{-1}$ for 30 minutes in channel no. 1 of the IBR-2 reactor of the Joint Institute for Nuclear Research (JINR, Dubna).\par
\indent  The chosen duration made it possible to avoid excessive irradiation and to perform coordinate measurements of tracks in $^{7}$Li $+ ^{4}$He events with a KSM microscope at a 90-fold magnification of the lens. Owing to a pronounced difference in the ionization of reaction products, the coordinates of the reaction vertex could be determined to a precision of 0.5 - 0.8 $\mu$m. The mean length of tracks of lithium nuclei was 3.1 $\pm$ 0.3 $\mu$m (RMS = 0.8 $\mu$m) at a mean thickness of 0.73 $\pm$ 0.02 $\mu$m (RMS = 0.05 $\mu$m); for the tracks of $^{4}$He nuclei, the respective values were 5.5 $\pm$ 0.5 $\mu$m (RMS = 1.1 $\mu$m) and 0.53 $\pm$ 0.01 $\mu$m (RMS = 0.04 $\mu$m). The directions of emission in pairs aren't collinear as a consequence of emission of $\gamma$-rays. A value of the average opening angle $\Theta$($^{7}$Li $+ ^{4}$He) is (148 $\pm$ 14)$^{\circ}$ (RMS = 35$^{\circ}$).  On the basis of these measurements, we obtained the angles at which the nuclei in the $^{7}$Li and $^{4}$He pairs fly apart (see Fig. 1). Because of photon emission, the directions of their motion were not collinear. The mean value of the angle $\Theta$($^{7}$Li $+ ^{4}$He) was 148$^{\circ}$ $\pm$ 14$^{\circ}$ (RMS = 35${^{\circ}$). In the distribution $\Theta$($^{7}$Li $+ ^{4}$He), there were several events in which $\Theta$($^{7}$Li $+ ^{4}$He) $<$ 90$^{\circ}$. Their origin might be due to visually indistinguishable alpha-particle scattering over the initial segment of their motion after the escape from the reaction vertex.\par
\begin{figure}
\includegraphics[width=0.45\textwidth]{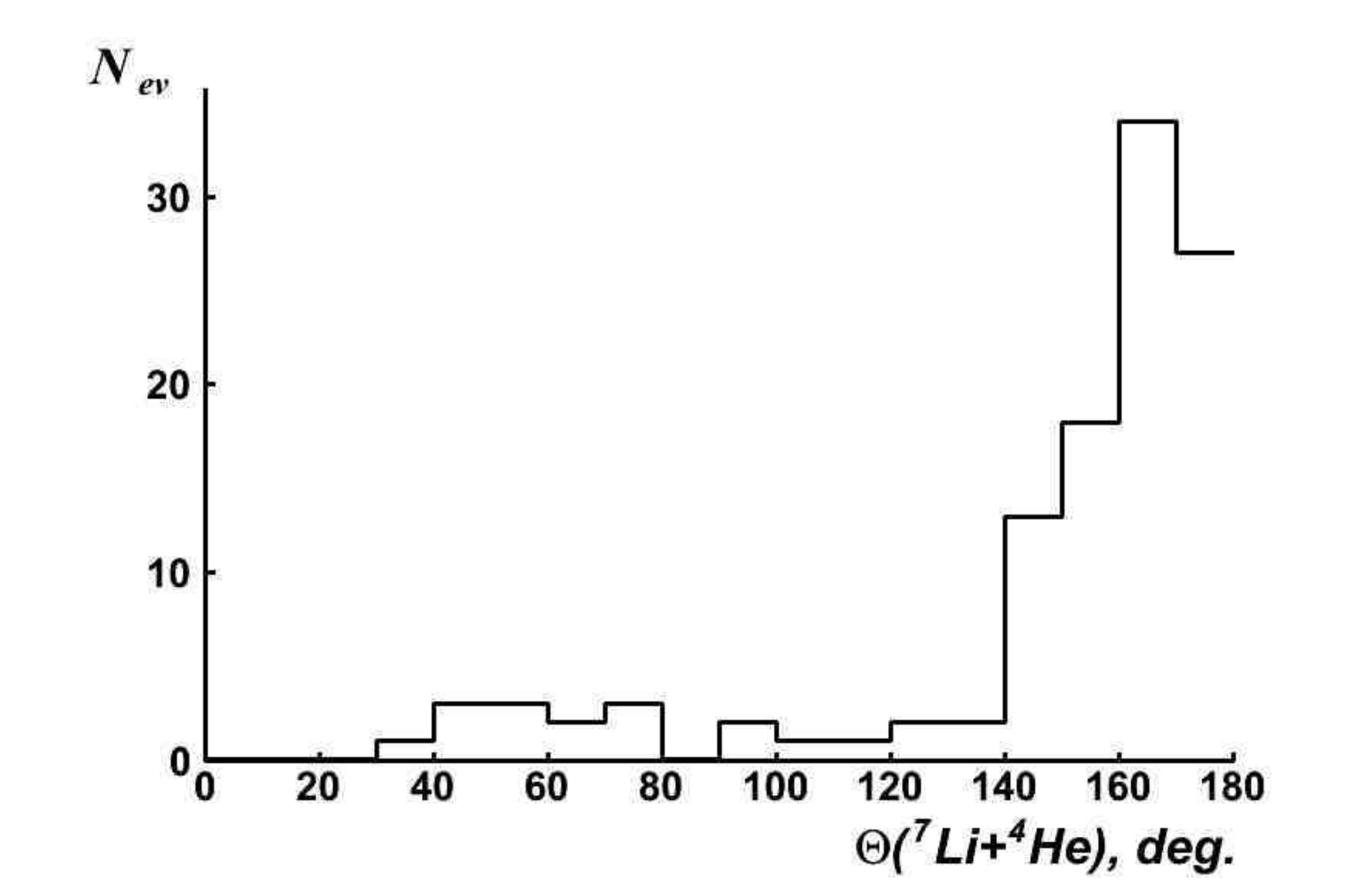}
\caption{Distribution of the angle $\Theta$($^{7}$Li $+ ^{4}$He)  at which the nuclei in $^{7}$Li and $^{4}$He pairs produced by thermal neutrons in 112 events of the reaction $n_{th} + ^{10}$B $\rightarrow ^{7}$Li $+ (\gamma) + ^{4}$He fly apart.}
\label{fig:1}
\end{figure} 

\indent The SRIÌ simulation code makes it possible to estimate the kinetic energy of nuclei on the basis of measurements of track lengths. Knowledge of the energies and emission angles permits determining
the distribution of the energy $Q(^{7}$Li $+ ^{4}$He) for pairs of $^{7}$Li and $^{4}$He nuclei  (see Fig. 2). The variable $Q$ is defined as the difference of the invariant mass of the final system, $M^{*}$, and the mass of the decaying nucleus, $M: Q = M^{*} - M$. Further, the square of of the mass $M^{*}$ is defined as the sum of all products of the fragment 4-momenta $P_{i,k}$; that is,
\begin{center}
$M^{*2} = (\Sigma P_{j})^{2} = \Sigma (P_{j} \cdot P_{k}) $
\end{center}
Its Lorentz-invariant character makes it possible to compare, on a unified basis, diverse data on nuclear reactions. Upon taking into account the energy carried away by emitted photon, the mean value of $Q(^{7}$Li $+ ^{4}$He), which is 2.4 $\pm$ 0.2 MeV (RMS = 0.8 MeV), complies with its expected counterpart.\par

\begin{figure}
\includegraphics[width=0.45\textwidth]{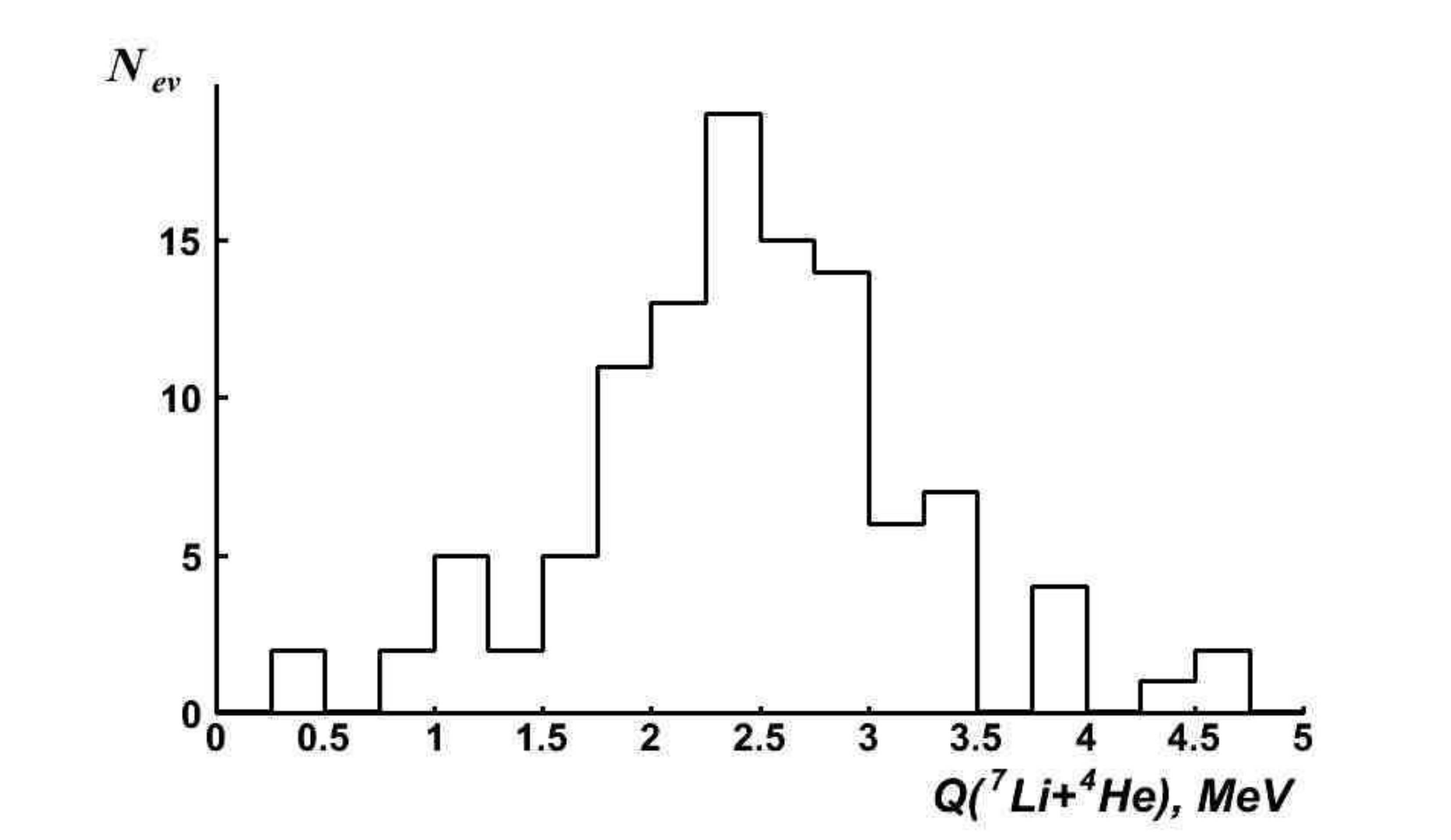}
\caption{Distribution of the energy Q($^{7}$Li$ + ^{4}$He) for pairs of $^{7}$Li and $^{4}$He nuclei produced by thermal neutrons in 112 events of the reaction $n_{th} + ^{10}$B $\rightarrow ^{7}$Li $+ (\gamma) + ^{4}$He.}
\label{fig:2}
\end{figure}

\indent The distribution of the angle $\Theta(\gamma + ^7$Li) between the photon-emission direction  calculated on the basis of the momentum-conservation law and the direction of emission of lithium nuclei exhibits an obvious anticorrelation (see Fig. 3). It is characterized by the mean value of $\Theta(\gamma + ^7$Li) = 128 $\pm$ 3$^{\circ}$ (RMS = 31$^{\circ}$) and by the coefficient 0.75 $\pm$ 0.07 characterizing asymmetry with respect to the angle of 90$^{\circ}$. In the case of $^4$He nuclei, the mean value of the angle $\Theta(\gamma + ^4$He) is 84 $\pm$ 4$^{\circ}$ (RMS = 40$^{\circ}$), while the coefficient of asymmetry if 0.14 $\pm$ 0.01.\par
\indent Thus, alpha-particle calibrations on the basis of the decays of $^{8}$He nuclei [5] and the breakup of $^{12}$C nuclei [6] were supplemented with applications in thermal-neutron beams and were extended to the $^7$Li nucleus. Calibration measurements on the basis of the reaction $n_{th} + ^6$Li $\rightarrow ^{4}$He $+ ^{3}$H in a track emulsion doped with lithium borate will become the next step along this line of research.\par

\begin{figure}
\includegraphics[width=0.45\textwidth]{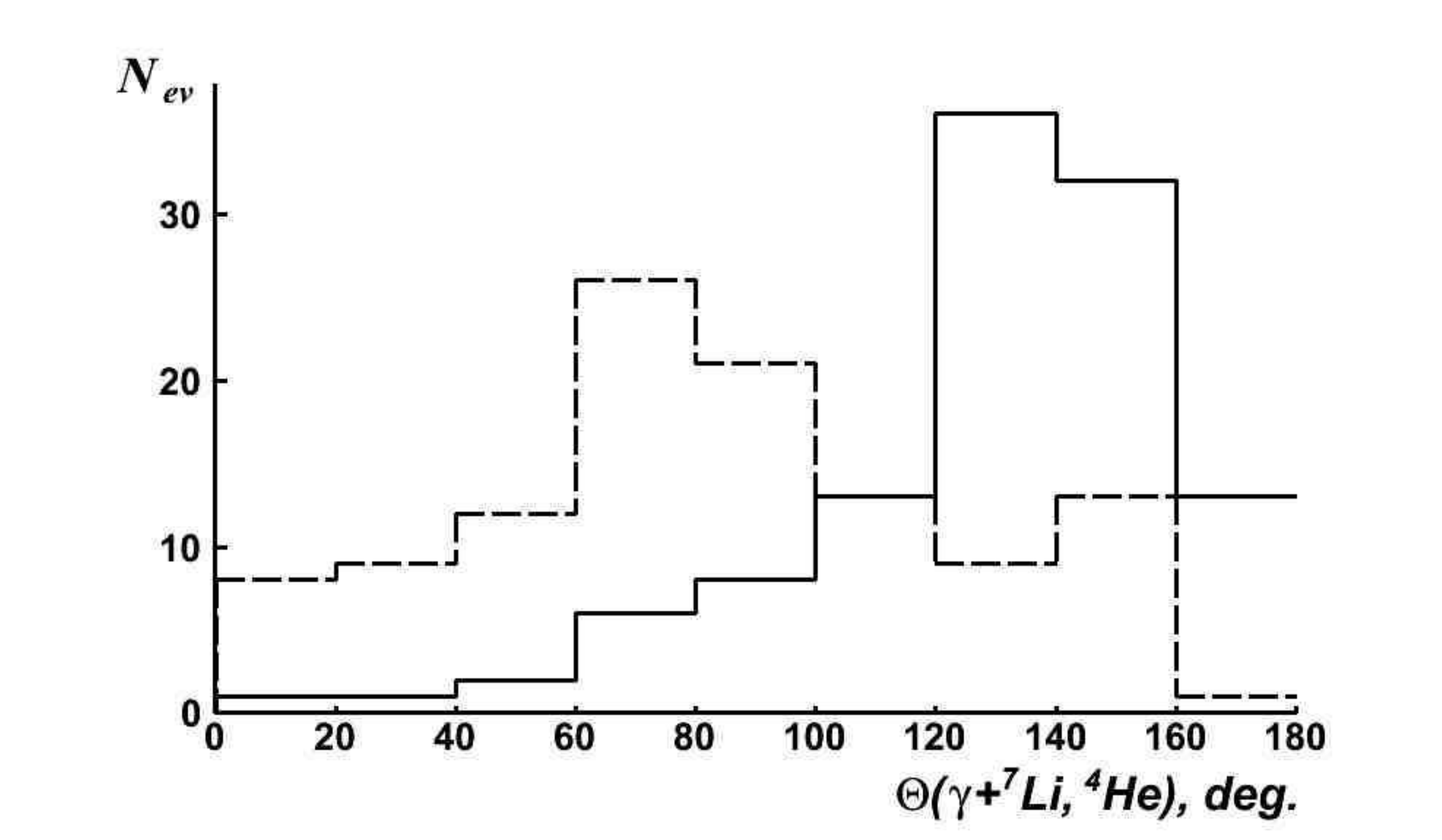}
\caption{Distribution of the angles $\Theta(\gamma + ^{7}$Li, $^{4}$He) between the calculated photon-emission direction and the emission direction of (solid-line histogram) $^{7}$Li and (dashed-line histogram) $^{4}$He nuclei produced by thermal neutrons in 112 events of the reaction $n_{th} + ^{10}$B $\rightarrow ^{7}$Li $+ (\gamma) + ^{4}$He.}
\label{fig:3}
\end{figure} 

\begin{center}
 EXPOSURE TO HEAVY IONS
\end{center}

\indent It is of interest to apply nuclear track emulsions in ternary-physics fission. The spontaneous fission of $^{252}$Cf nuclei or $^{235}$U fission induced by thermal neutrons is likely to be a subject of searches for molecule-like nuclear systems. The emission of respective fragments may prove to be collinear. In the decay of a three-body system, one of the heavy fragments may entrain the light fragment. A track emulsion will make it possible to study correlations at small angles for fragments of collinear ternary fission. It is assumed that the track emulsion used is actuated by fission fragments upon coming into contact with the film onto which the isotope under study is deposited. The track emulsion irradiated with a $^{252}$Cf source yielding primarily alpha particles and, with a probability of 6\%, spontaneous-fission fragments and also, for the sake of comparison, with a $^{241}$Am source  yielding only alpha particles is being presently analyzed.\par

\begin{figure}
\includegraphics[width=0.45\textwidth]{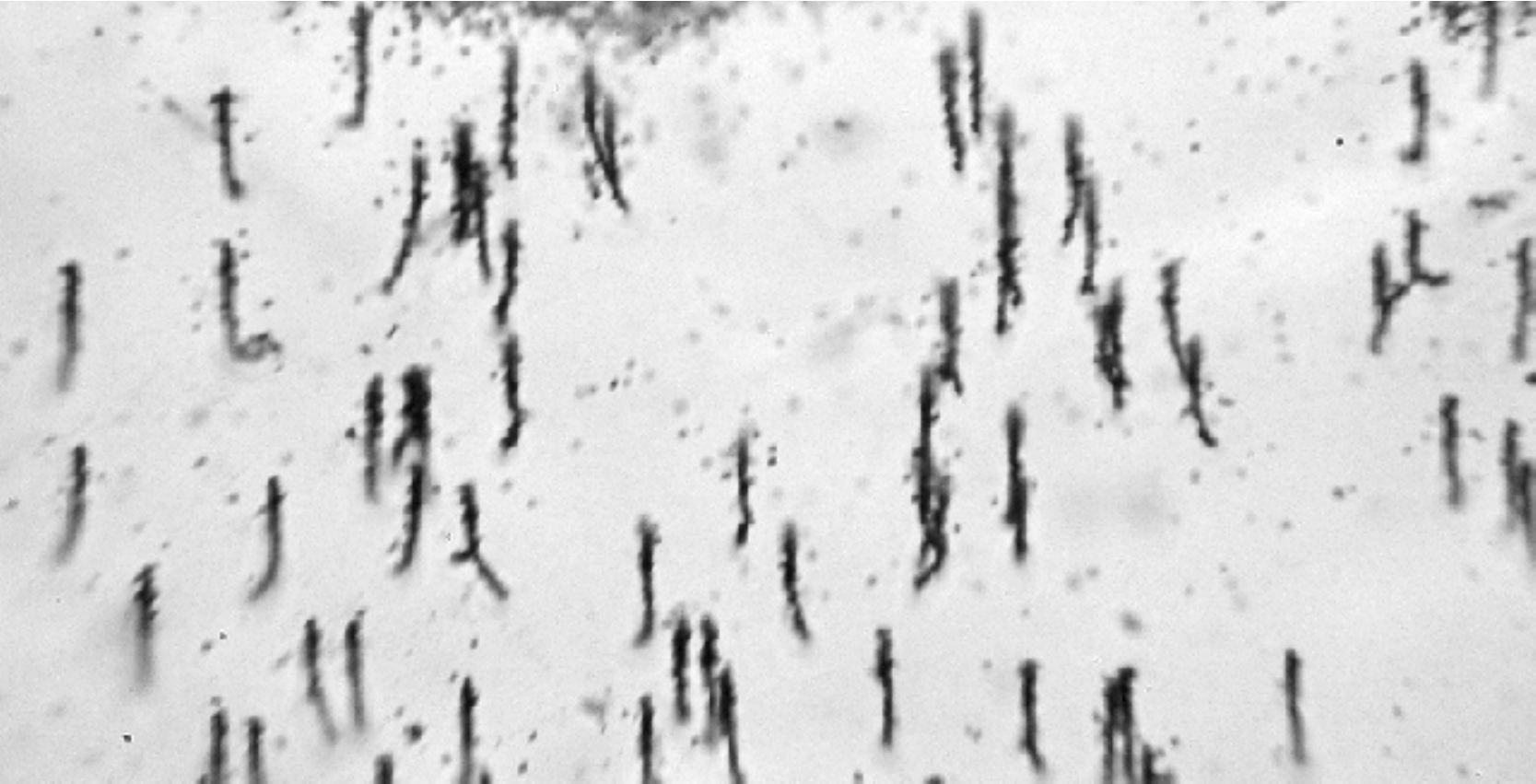}
\caption{Closeup photography of the track-emulsion segment featuring traces of stopped xenon ions (the ions enter the track-emulsion layer from above) at a 90-fold magnification.}
\label{fig:4}
\end{figure} 

\indent It is necessary to calibrate of ranges and estimation of the angular distribution for the maximally wider variety of heavy ions of known energy that were implanted in the track emulsion. It is of importance to extend the energy calibration to the region of energies below the Coulomb barrier for nuclear reactions. The experience of the range-based spectroscopy of heavy nuclei will be of use in searches for hypothetical dark-matter particles.\par
\indent The track emulsion used was irradiated with $^{86}$Kr$^{+17}$ and $^{132}$Xe$^{+26}$ accelerated to an energy of about 1.2 MeV per nucleon at the ITS-100 cyclotron of the G.N. Flerov Laboratory of Nuclear Reactions at JINR. Since the energy of these ions is low, the track emulsion was exposed without a backing paper. Therefore, the fixation of track-emulsion plates in the irradiation chamber was performed under ordinary illumination in a photographic processing laboratory. Over five seconds of irradiation, the density of tracks was $10^5$ to $10^6$nucl.$/$cm$^{2}$. Track-emulsion layers of area 9$\times$12 cm and thickness 161 $\pm$ 10 $\mu$m for krypton and 119 $\pm$ 3 $\mu$m for xenon were installed at an angle 45$^{\circ}$ with respect to the beam axis. This made it possible to observe the stopping of ions. Figure 4 shows a closeup photography of the track-emulsion segment carrying traces of xenon. This photography was made with an MBI-9 microscope at a 90-fold magnification of the lens.\par
\indent The lengths of tracks of ions stopped in track-emulsion layers without undergoing scattering were measured with a KSM microscope at a 90-fold magnification. The mean length of tracks without scattering proved to be 14.3 $\pm$ 0.15 $\mu$m (RMS = 0.9 $\mu$m) for krypton ions and 17.5 $\pm$ 0.1 $\mu$m (RMS = 1.0 $\mu$m) for Xe ions. These values are close to their counterparts calculated on the basis of the SRIM model: 18.5 $\pm$ 1.3 $\mu$m (RMS = 1.3 $\mu$m) for krypton and 20.1 $\pm$ 2.2 $\mu$m (RMS = 1.3 $\mu$m) for xenon. On the basis of spline interpolation of the results of range–energy calculations, measurements of the ion-track lengths make it possible to estimate the kinetic energies of ions by using the SRIM model. Its mean value if 0.74 $\pm$ 0.01 MeV per nucleon (RMS = 0.1 MeV per nucleon) for krypton ions and 0.92 $\pm$ 0.01 MeV per nucleon (RMS = 0.1 MeV per nucleon) for xenon ions. The measured mean values proved to be somewhat smaller than those that we expected, which indicates that the simulation parameters call for refinements. The mean angle at which ions enter the track-emulsion layer is 43.8$\circ \pm$ 0.6$\circ$ (RMS = 4$\circ$) for krypton and 44.7$\circ \pm$ 0.6$\circ$ (RMS = 4$\circ$) for xenon. This corresponds to the angle of orientation of the plate containing a track emulsion with respect to the beam axis.\par
\indent Many of the primary tracks end in incurvations and vees, which result from scattering on nuclei forming track emulsions. In the course of moderation before scattering, the ion energy decreases to values one order of magnitude lower than the Coulomb barrier. On the basis of detailed coordinate measurements for vees, it is assumed to identify observed recoil nuclei and to extend the investigation of the energy resolution to extremely low energies. Thereby, one can reconstruct the kinematics of scattering and subject the track-emulsion resolution to a new test. This aspect is of importance for the calibration of track emulsions with a submicron resolution for seeking dark-matter particles.\par

\newpage
\begin{center}
 EXPOSURE TO MUONS
\end{center}

\indent The deep-inelastic scattering of ultrarelativistic muons is a commonly recognized means for studying the parton structure of nucleons and nuclei. The irradiation of track emulsions with these particles makes it possible to study concurrently the multifragmentation of nuclei under the effect of a purely electromagnetic probe. Multiphoton exchanges or transitions of virtual photons to vector mesons may serve as a fragmentation mechanism. At CERN, a track-emulsion sample was exposed to 160-GeV muons. Previously, such irradiations of track emulsions were not performed. The objective of the exposure described here was to study experimental loads in the vicinity of the beam axis and to assess preliminarily the character of muon interactions.\par

\begin{table}
\caption{Distribution of the numbers of strongly ionizing $b-$ and $g$ particles,  $N_b$ and  $N_g$, in stars formed by muons for $N_b + N_g > 13$}
\begin{center}
\begin{tabular}{c|c} \hline
~  $N_b$  ~&~  $N_g$  ~\\ \hline
~  13  ~&~  1  ~\\ 
~  11  ~&~  4  ~\\ 
~  11  ~&~  6  ~\\ 
~  12  ~&~  6  ~\\ 
~  11  ~&~  5  ~\\ 
~  13  ~&~  2  ~\\ 
~  12  ~&~  3  ~\\ 
~  5  ~&~  10  ~\\ 
~  11  ~&~  7  ~\\ 
~  6  ~&~  9  ~\\ \hline
\end{tabular}
\end{center}
\end{table}
\indent The samples under study were placed in front of the target of the COMPASS experiment at a distance of about 25 cm from the beam axis (halo), where the intensity amounted to about $10^{6}$ particles per centimeter squared per cycle. The track-emulsion samples 9 $\times$ 12 cm in area and about 100 $\mu$m in thickness were oriented both along and across the beam. A nine-hour irradiation in the case of the transverse orientation proved to be the most favorable for our analysis. So long-term an irradiation was possible owing to the smallness of the cross section for muon interaction and to a small effect of beam ionization in relation to the longitudinal arrangements of layers. The duration of the irradiation run was constrained for fear of overloads with tracks from interactions in the glass substrate. In principle, this duration could be increased by two orders of magnitude without causing complications for the ensuing analysis.\par
\indent Scanning led to finding, in irradiated track emulsions, about 300 stars containing not less than three target fragments. The topology of stars was determined by the number of strongly ionizing $b$ and $g$ particles ($N_b$ and $N_g$, respectively). Figure 5 shows the distribution of $N_b$. In the brightest ten events, the total number of the respective tracks, $N_b$ and $N_g$ ($N_h$), proved to be not less than 14 (see table). Although the solid angle within which tracks could be observed was limited, the formation of high-multiplicity stars involving almost one-half of the charge of heavy nuclei in the composition of the emulsion could be proven.\par

\begin{figure}
\includegraphics[width=0.45\textwidth]{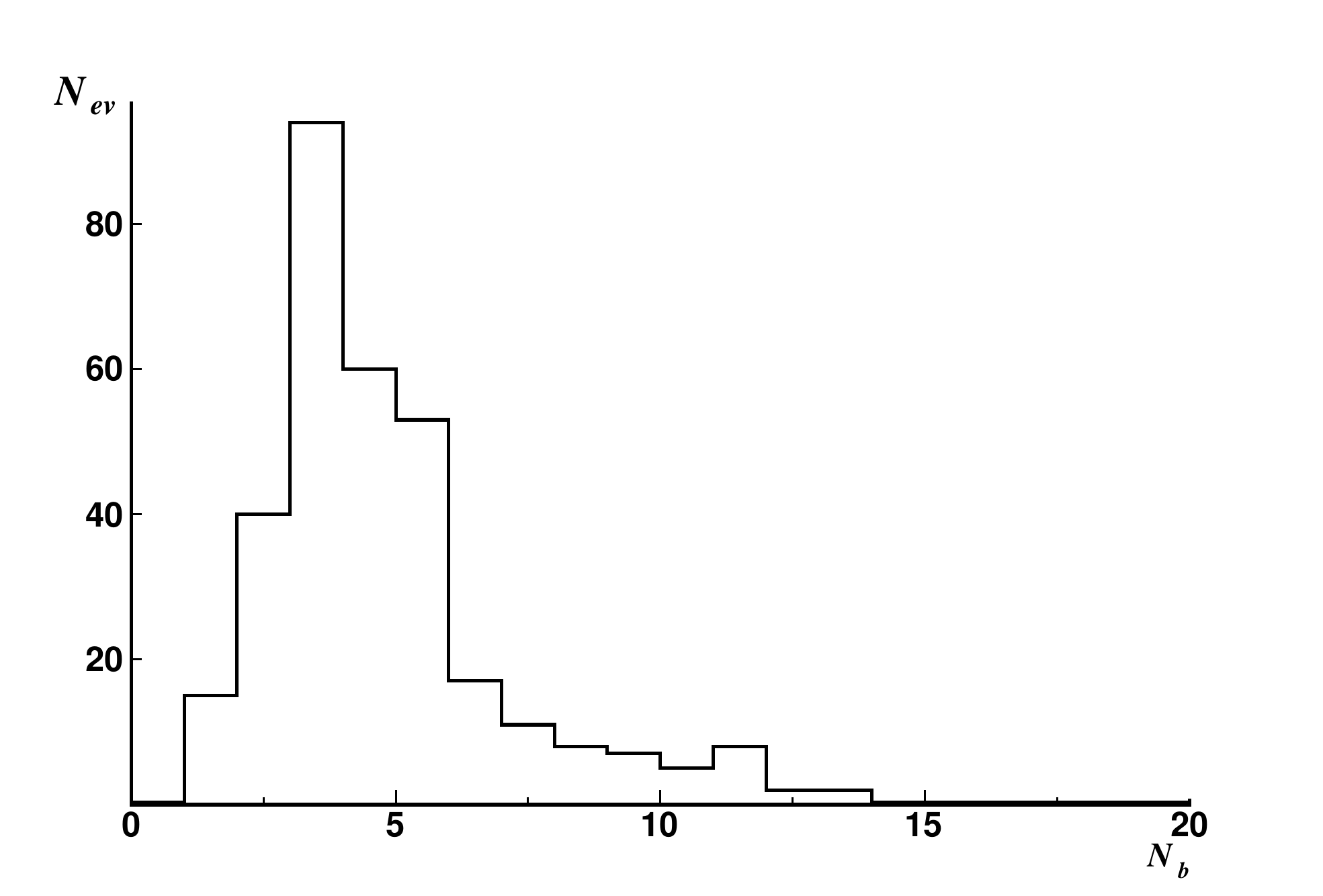}
\caption{Distribution of nuclear stars with respect to the number of strongly ionizing $b$ particles, $N_{b}$, in a track-emulsion sample irradiated with muons.}
\label{fig:5}
\end{figure} 

\indent Seventy-two stars containing only triples of $b$ particles stopped in the track emulsion were associated with the breakup process $^{12}$Ñ $\rightarrow$ 3$\alpha$. Alpha-particle ranges and spatial emission angles were determined on the basis of coordinate measurements for tracks. The mean alpha-particle range was 23.1 $\pm$ 0.6 $\mu$m (RMS =8.4 $\mu$m).  The alpha-particle energy was estimated on the basis of the SRIM model. Its mean value proved to be 5.3 $\pm$ 0.1 MeV (RMS =1.3 MeV). Figure 6 total-longitudinal-momentum ($P_{z}$) distribution of alpha-particle triads. As might have been expected, it is basically concentrated in the region of $P_{z} >$ 0, this being indicative of the arrival direction of beam particles. A moderately small number of $P_{z} <$ 0 events stem from the contribution of background particles produced in the backward direction on target or setup materials.\par

\indent The fact that the interpretation of this group of events is unambiguous makes it possible to assess the character of their production on the basis of the total transverse momentum $P_T$ of alpha-particle triads. The distribution of $P_T$ (see Fig. 7) is characterized by the mean value of 241 $\pm$ 28 MeV$/c$ (RMS = 123 MeV$/c$). It is described by the Rayleigh distribution at the parameter value of 190 $\pm$ 13 MeV$/c$. These parameter values are typical of the nuclear diffraction interaction. In the case of purely electromagnetic exchange, the $P_T$ distribution would be concentrated in the range of $P_T <$ 100 MeV$/c$. It is useful to compare this distribution of $P_T$ with the substantially narrower distribution of $P_{T}$ for the reaction $n$(14.1 MeV)$ + ^{12}$C $\rightarrow$ 3$\alpha + n$ (see Fig. 5  in[3]). The latter is characterized by the mean value of 69 $\pm$ 4 MeV$/c$ (RMS = 38 MeV$/c$ ) and by the Rayleigh distribution parameter value of 55 $\pm$ 28 MeV$/c$.\par

\begin{figure}
\includegraphics[width=0.45\textwidth]{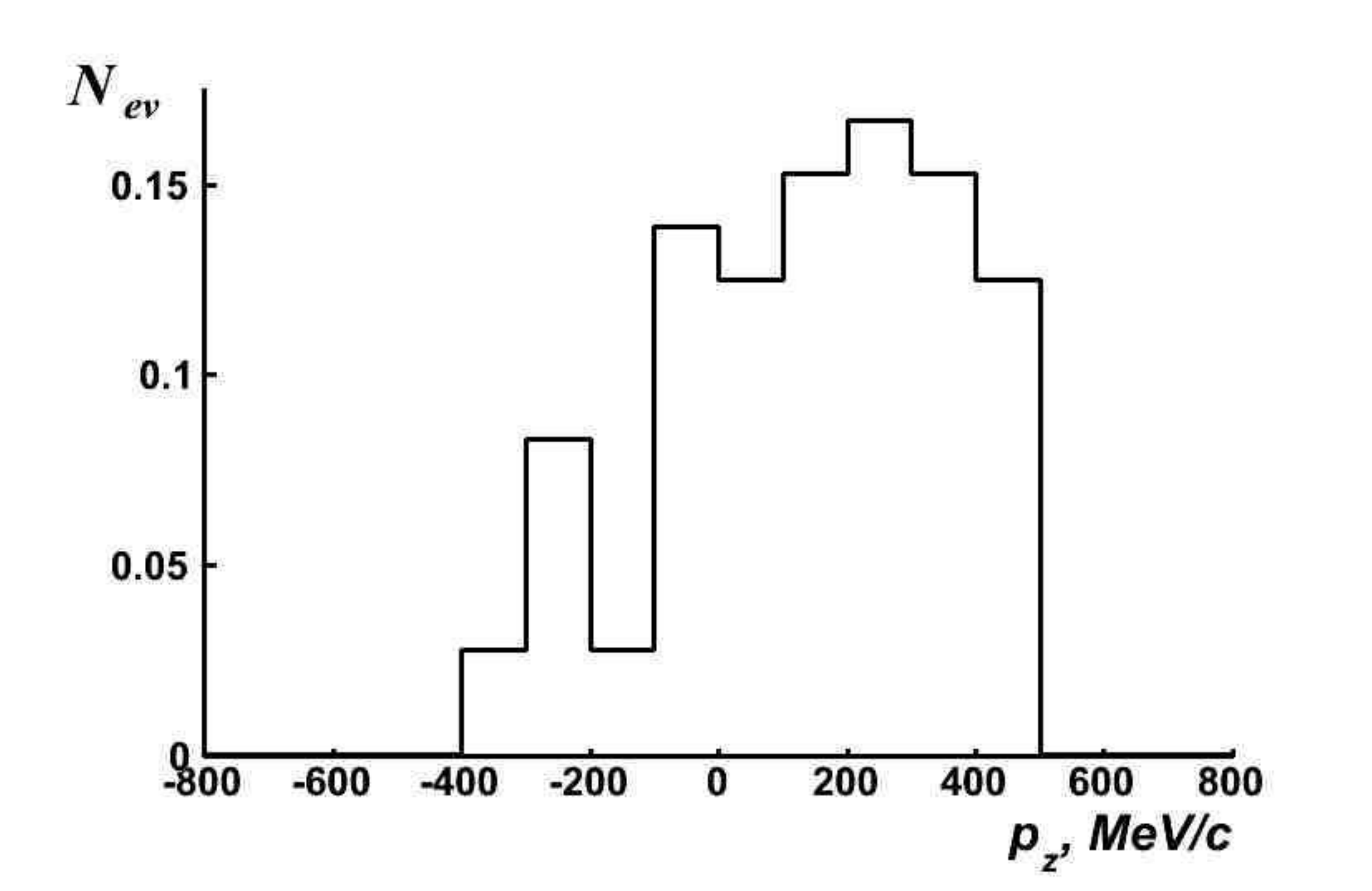}
\caption{ Total-longitudinal-momentum ($P_{z}$) distribution of 72 triads of alpha particles in a track emulsion irradiated with muons (normalization to the number of events).}
\label{fig:6}
\end{figure} 
\begin{figure}
\includegraphics[width=0.45\textwidth]{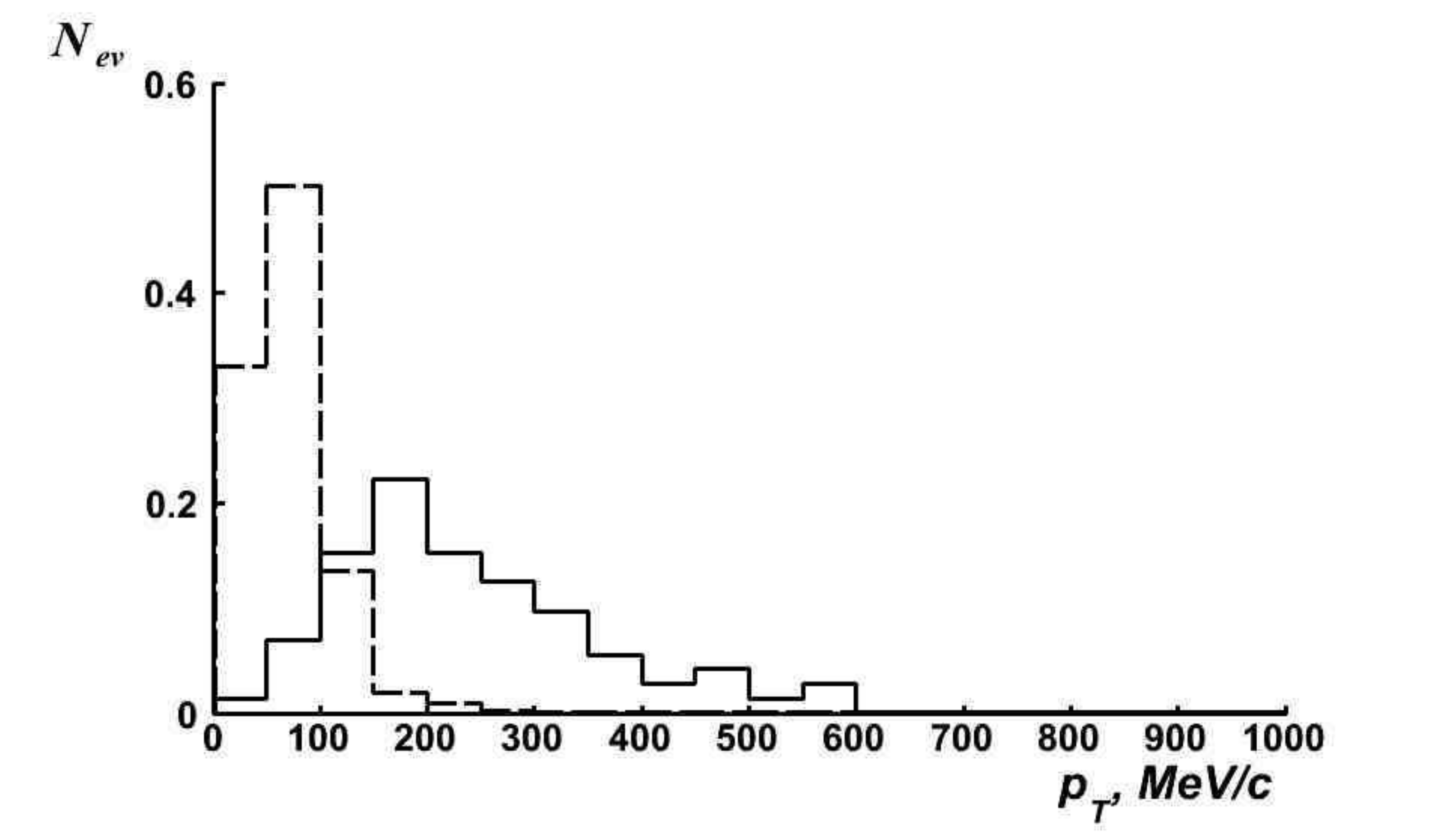}
\caption{Total-transverse-momentum ($P_{T}$) distribution of (solid-line histogram) 72 triads of alpha particles in a track emulsion irradiated with muons and (dashed-line histogram) 400 triads of alpha particles from the reaction $n$(14.1 MeV) $+ ^{12}$C $\rightarrow$ 3$\alpha + n$ [3] (normalization to the number of events).}
\label{fig:7}
\end{figure} 

\indent The distribution of the total energy of alpha-particle triads, $Q_{3\alpha}$, in Fig. 8 is  substantially broader than that in the case of the reaction $n$(14.1 MeV) $+ ^{12}$C $\rightarrow$ 3$\alpha$ [6], the latter revealing distinctly the clustering features of the $^{12}$C nucleus. In the case being considered, the distribution of $Q_{3\alpha}$ is concentrated above the alpha-cluster levels of excitation of the $^{12}$C nucleus. The distribution of the energy of alpha-particle pairs, $Q_{2\alpha}$ , in Fig. 9 does not reveal any similarity in the spectra for the exposures being considered either. In the exposure to muons, there is virtually no signal from the decays of the ground state $^{8}$Be$_{g.s.}$ in the range of $Q_{2\alpha} <$ 200 keV (see inset in Fig. 9), which manifest themselves as "narrow" alpha-particle pairs [6]. The distribution of $Q_{2\alpha}$ does not exhibit a peak from the decays of the first excited state $^{8}$Be$^{2+}$ at 3 MeV. Moreover, this distribution of $Q_{2\alpha}$ proves to be substantially broader than that in the case of the reaction $n$(14.1 MeV) $+ ^{12}$C $\rightarrow$ 3$\alpha$ [6].

\indent By and large, the $P_{T}, Q_{3\alpha}$ and $Q_{2\alpha}$ distributions for the irradiation of nuclear track emulsions with muons are indicative of a hard character of the process without manifestations of the well-known structural features of the $^{12}$C nucleus, including the formation of alpha-particle triads in the continuum region. We emphasize that the contribution of $^{12}$C breakup with a threshold of 7.36 MeV should have inevitably manifested itself in the channel being discussed ($N_{b} = 3$ and $N_{g} = 0$). However, the circumstance that it is the nuclear diffraction mechanism rather than the soft electromagnetic mechanism that manifests itself for this channel, which possesses the minimum threshold, seems unexpected and deserves a theoretical analysis. The corroboration of this conclusion is of importance for interpreting not only multifragmentation under the effect of ultrarelativistic muons. It may also serve as a basis for interpreting the multifragmentation of relativistic nuclei in peripheral interactions not leading to the formation of target fragments (white stars).
\indent These observations, which are of a preliminary character, indicate that a full-scale investigation of a complete muon-induced disintegration of nuclei on the basis of multilayered assemblies from thick layers of substrate-free nuclear track emulsion is highly promising. In order to interpret reliably data obtained upon exposing track-emulsion layers to muons, it is necessary to test the hadron-background level at the places where the emulsion layers were irradiated. Data from such an irradiation could be used in planning experiments based on silicon detectors or on a time-projection chamber. Investigation of the muon-induced fragmentation of nuclei is of practical interest for developing approaches to the separation of muons and pions on the basis of the distinctions between the stars created by them. Moreover, this is useful for testing models of physics processes for lepton–nucleus colliders.

\begin{figure}
\includegraphics[width=0.45\textwidth]{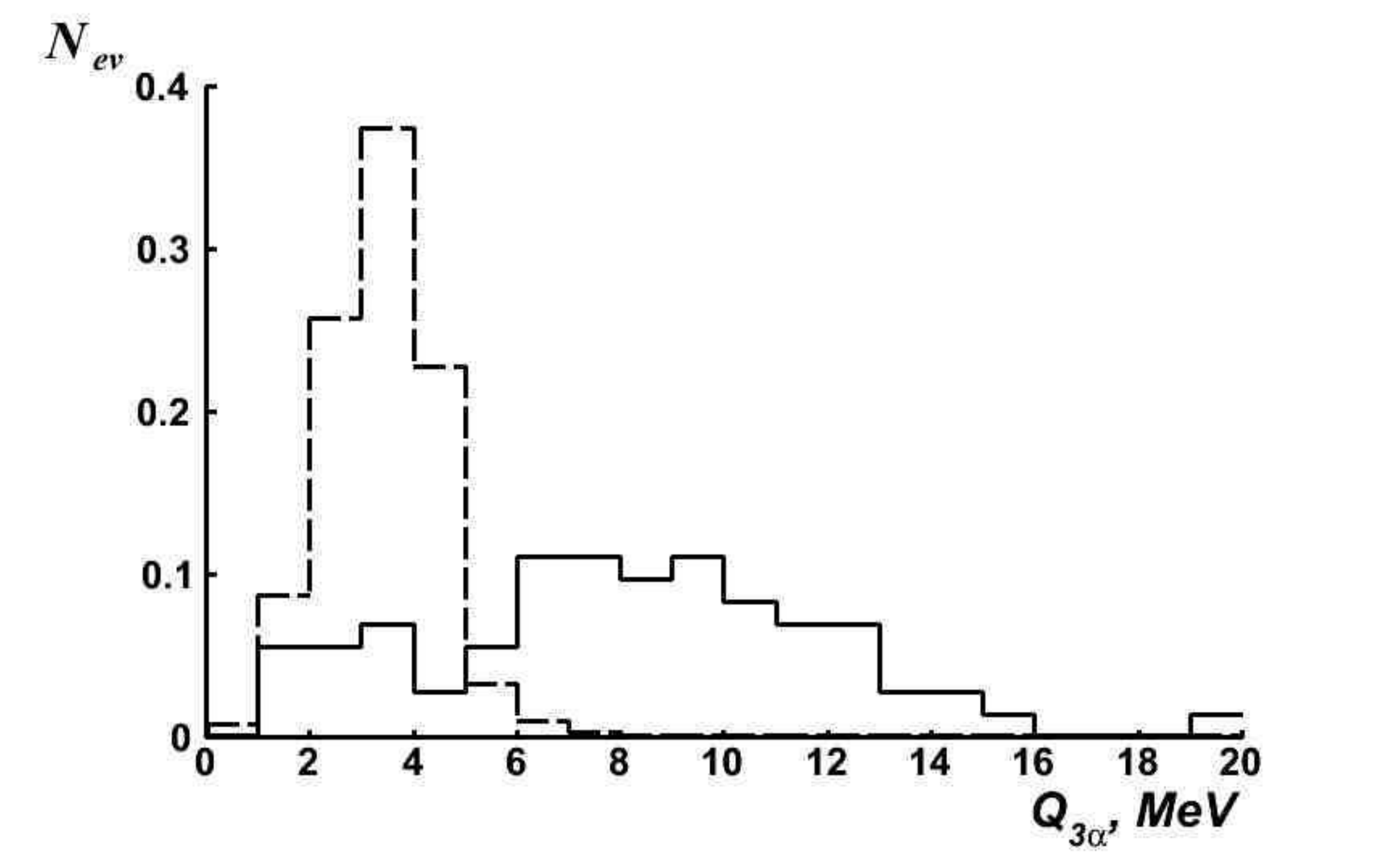}
\caption{Distribution of the total energy of alpha-particle triads, $Q_{3\alpha}$, in a track emulsion irradiated with (solid-line histogram) muons and (dashed-line histogram) neutrons in the reaction $n$(14.1 MeV) $+ ^{12}$C $\rightarrow$ 3$\alpha + n$ [3] (normalization to the number of events).}
\label{fig:8}
\end{figure} 
\begin{figure}
\includegraphics[width=0.45\textwidth]{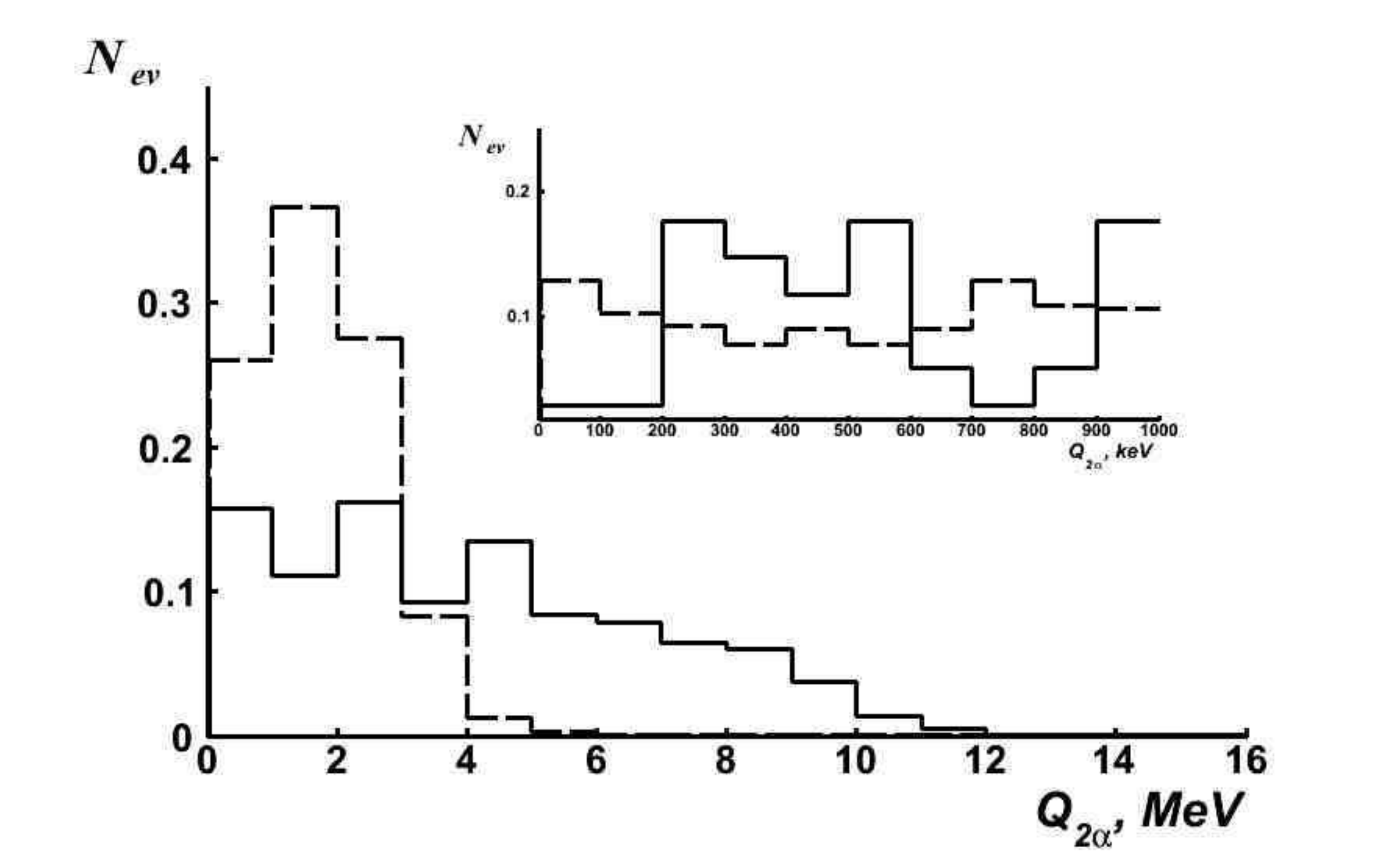}
\caption{Distribution of the energy of alpha-particle pairs, $Q_{2\alpha}$, in a track emulsion irradiated with (solid-line histogram) muons and (dashed-line histogram) neutrons in the reaction $n$(14.1 MeV) $+ ^{12}$C $\rightarrow$ 3$\alpha + n$ [3] (normalization to the respective number of pairs).}
\label{fig:9}
\end{figure} 

\begin{center}
ACKNOWLEDGMENTS
\end{center}

\indent We are grateful to our colleagues whose contribution made it possible to perform a series of irradiation runs in 2011–2013. The use of new track-emulsion samples became possible owing to the extensive support of O.I.~Orurk (Slavich Company JSC), coordinator of this project, and to the creative work of Yu.A.~Berezkina, A.V.~Kuznetsov, and L.V.~Balabanova, members of the staff of the MICRON production unit of this company (Slavich Company, Pereslavl-Zalessky). The generous help of A.S.~Mikhailov (Moscow Institute of Kino and Video) in developing strategies for the reproduction of track-emulsion technologies is gratefully acknowledged.
\indent The implementation of the irradiation of track-emulsion samples was due to efforts of members of the JINR staff: S.B.~Borzakov in the exposures to a thermal-neutron beam from the IBR-2 reactor; O.M.~Ivanov in the exposures at the ITS-100 cyclotron; and O.P.~Gavrishchuk, G.V.~Meshcheryakov, and A.S.~Nagaitsev in the exposures to a muon beam at CERN. We are also indebted to A.I.~Malakhov from JINR and to N.G.~Polukhina and S.P.~Kharlamov from Lebedev Physical Institute (Russian Academy of Sciences) for their support and constructive criticism in discussing our results.
\indent This work was supported by the Russian Foundation for Basic Research (project no. 12-02-00067) and by grants from the plenipotentiaries of Bulgaria, Romania, and Czech Republic at JINR.\par
\indent The JINR staff members provided the NTE samples irradiation of: S.~B.~Borzakov – at the IBR-2 thermal neutron beam, O.~M.~Ivanov - at the cyclotron IC-100, O.~P.~Gavrishchuk, G.~V.~Mescheryakov and A.~S.~Nagaytsev – at the $\mu$-meson beam at CERN. This work was supported by grant from the Russian Foundation for Basic Research 12-02-00067 and grants plenipotentiary representatives of Bulgaria, Romania and the Czech Republic at JINR.\par


\begin{thebibliography}{99}

\bibitem{01} P.~I.~Zarubin, Lect. Notes in Phys., 875, 51(2013) Springer Int. Publ.; arXiv:1309.4881 [nucl-ex].\par
\bibitem{02} K.~Z.~Mamatkulov et al., Phys. At. Nucl. 76, 1224(2013); arXiv:1309.4241 [nucl-ex].\par
\bibitem{03} R.~R.~Kattabekov et al., Phys. At. Nucl. 76, 1219(2013);  arXiv:1310.2080 [nucl-ex].\par
\bibitem{04} N.~K.~Kornegrutsa et al., Phys. At. Nucl 76, add. issue, 84(2013).\par
\bibitem{05} D.~A.~Artemenkov et al., Phys. of Part. and Nucl. Lett., 10, 415 (2013); arXiv:1309.4808.\par
\bibitem{06} R.~R.~Kattabekov et al., Phys. At. Nucl., add. issue, 88(2013).\par
\bibitem{07} CERN Courier 54, 1, 42(2014); http://www.spaces\\cience.ro/wnte2013/\par
\bibitem{08} The BECQUEREL Project, http://becquerel.jinr.ru/\par
\bibitem{09} Slavich Company JSC,  www.slavich.ru, www.new\\slavich.com.\par
\bibitem{10} J.~F.~Ziegler, J.~P.~Biersack and M.~D.~Ziegler. SRIM - The Stopping and Range of Ions in Matter  2008, ISBN \\
0-9654207-1-X., SRIM Co; http://srim.org/.\par
\bibitem{11} D.~V.~Kamanin and Y.~V.~Pyatkov, Lect. Notes in Phys., 875, 183(2013) Springer Int. Publ.\par
\end{thebibliography}
\end{document}